\documentclass[12pt]{article}
\usepackage{graphicx}

\def\hybrid{\topmargin 0pt      \oddsidemargin 0pt
        \headheight 0pt \headsep 0pt
       \voffset-1cm
        \textwidth 6.25in       
       \textheight 9.5in       
        \marginparwidth 0.0in
        \parskip 5pt plus 1pt   \jot = 1.5ex}
\catcode`\@=11
\def\marginnote#1{}

\newcount\hour
\newcount\minute
\newtoks\amorpm
\hour=\time\divide\hour by60
\minute=\time{\multiply\hour by60 \global\advance\minute by-\hour}
\edef\standardtime{{\ifnum\hour<12 \global\amorpm={am}%
        \else\global\amorpm={pm}\advance\hour by-12 \fi
        \ifnum\hour=0 \hour=12 \fi
        \number\hour:\ifnum\minute<10 0\fi\number\minute\the\amorpm}}
\edef\militarytime{\number\hour:\ifnum\minute<10 0\fi\number\minute}

\def\draftlabel#1{{\@bsphack\if@filesw {\let\thepage\relax
   \xdef\@gtempa{\write\@auxout{\string
      \newlabel{#1}{{\@currentlabel}{\thepage}}}}}\@gtempa
   \if@nobreak \ifvmode\nobreak\fi\fi\fi\@esphack}
        \gdef\@eqnlabel{#1}}
\def\@eqnlabel{}
\def\@vacuum{}
\def\draftmarginnote#1{\marginpar{\raggedright\scriptsize\tt#1}}

\def\draftlabel#1{{\@bsphack\if@filesw {\let\thepage\relax
   \xdef\@gtempa{\write\@auxout{\string
      \newlabel{#1}{{\@currentlabel}{\thepage}}}}}\@gtempa
   \if@nobreak \ifvmode\nobreak\fi\fi\fi\@esphack}
        \gdef\@eqnlabel{#1}}
\def\@eqnlabel{}
\def\@vacuum{}
\def\draftmarginnote#1{\marginpar{\raggedright\scriptsize\tt#1}}

\def\draft{\oddsidemargin -.5truein
        \def\@oddfoot{\sl preliminary draft \hfil
        \rm\thepage\hfil\sl\today\quad\militarytime}
        \let\@evenfoot\@oddfoot \overfullrule 3pt
        \let\label=\draftlabel
        \let\marginnote=\draftmarginnote
   \def\@eqnnum{(\theequation)\rlap{\kern\marginparsep\tt\@eqnlabel}%
\global\let\@eqnlabel\@vacuum}  }

\def\numberbysection{\@addtoreset{equation}{section}
        \def\theequation{\thesection.\arabic{equation}}}

\def\underline#1{\relax\ifmmode\@@underline#1\else
        $\@@underline{\hbox{#1}}$\relax\fi}

\def\titlepage{\@restonecolfalse\if@twocolumn\@restonecoltrue\onecolumn
     \else \newpage \fi \thispagestyle{empty}\c@page\z@
        \def\thefootnote{\fnsymbol{footnote}} }

\def\endtitlepage{\if@restonecol\twocolumn \else  \fi
        \def\thefootnote{\arabic{footnote}}
        \setcounter{footnote}{0}}  
\relax

\numberbysection
\hybrid

\newfont{\Bbb}{msbm10 scaled 1\@ptsize00}
\newfont{\Bbbb}{msbm7 scaled 1\@ptsize00}

\newcommand{\DDD}{\raise-1pt\hbox{$\mbox{\Bbbb D}$}}

\newcommand{\UUU}{\raise-1pt\hbox{$\mbox{\Bbbb U}$}}

\newcommand{\ZZ}{\mbox{\Bbb Z}}
\newcommand{\z}{\raise-1pt\hbox{$\mbox{\Bbbb Z}$}}

\def\beq{\begin{equation}}
\def\eeq{\end{equation}}
\def\p{\partial}

\def\res{\mathop{\hbox{res}}\limits}

\newtheorem{lemma-definition}{Lemma-Definition}[section]

\begin{document}

\begin{titlepage}

\title{Kadomtsev-Petviashvili hierarchies of types B and C}

\author{A.~Zabrodin
\thanks{
Skolkovo Institute of Science and Technology, 143026, Moscow, Russia and
Institute of Biochemical Physics, Kosygina str. 4, 119334, Moscow, Russia and
ITEP NRC KI, 25
B.Cheremushkinskaya, Moscow 117218, Russia;
e-mail: zabrodin@itep.ru}}

\date{February 2021}
\maketitle

\vspace{-0.8cm}

\begin{center}

{\it Dedicated to the memory of S.M. Natanzon}

\end{center}

\vspace{-7cm} \centerline{ \hfill ITEP-TH-04/21}\vspace{7cm}

\begin{abstract}

This is a short review of the Kadomtsev-Petviashvili
hierarchies of types B and C. The main objects are the $L$-operator, the wave operator, 
the auxiliary linear 
problems for the wave function, the bilinear identity for the wave function
and the tau-function. All of them are discussed in the paper.
The connections with the usual Kadomtsev-Petviashvili
hierarchy (of the type A) are clarified. Examples of soliton solutions
and the dispersionless limit of the hierarchies are also considered.

\end{abstract}

\end{titlepage}

\vspace{5mm}

%

\tableofcontents

\vspace{5mm} 

\section{Introduction}

In the papers \cite{DJKM83,DJKM82,DJKM81} infinite integrable hierarchies of partial differential
equations with ${\rm O}\, (\infty)$ and 
${\rm Sp}\, (\infty )$ symmetry were introduced. They can be called the 
Kadomtsev-Petviashvili hierarchies of type B (BKP) and C (CKP).
The BKP (respectively, CKP) hierarchy was also discussed in \cite{JimboMiwa,LW99,Tu07,DM-H09}.
(respectively, \cite{DM-H09,CW13,CH14,LOS12,KZ20}). 

As is pointed out in \cite{DJKM81}, the general solutions to the BKP
and CKP hierarchies depend on 
functional parameters in two variables. 
In a certain sense to be clarified below
in the main text these hierarchies can be regarded as restrictions of the well known
Kadomtsev-Petviashvili (KP) hierarchy. In a nutshell, this can be made more precise
as follows. Let ${\cal X}_{\rm KP}$ be the moduli space of solutions to the KP hierarchy
(according to Segal and Wilson, 
it is an infinite dimensional Grassmann manifold). The modili spaces of solutions
to the BKP and CKP hierarchies are submanifolds of ${\cal X}_{\rm KP}$:
${\cal X}_{\rm BKP}\subset {\cal X}_{\rm KP}$, 
${\cal X}_{\rm CKP}\subset {\cal X}_{\rm KP}$, and the ``even'' time evolution 
(i.e., the evolution with respect to the times $t_{2k}$, $k\geq 1$) is frozen.

This paper is a short review of the BKP and CKP hierarchies. We discuss the main
objects and notions related to them: the $L$-operator, the wave operator, 
the auxiliary linear 
problems for the wave function, the bilinear identity for the wave function
and the tau-function. The tau-function satisfies certain equations (the Hirota equations) 
which are bilinear
in the BKP case and have a more complicated structure in the CKP case.
The connection between tau-functions of the KP, BKP and CKP
hierarchies is clarified. 

As examples of solutions, we give 
explicit formulas for soliton solutions. BKP and CKP $N$-solitons are specializations of
$2N$ soliton solutions of the KP hierarchy. As is known, soliton solutions are degenerations
of more general quasi-periodic (algebro-geometric) solutions. 
According to the Krichever's construction \cite{Krichever77a}, 
any smooth algebraic curve with some 
additional data provides a quasi-periodic solution.
The quasi-periodic 
solutions of the BKP hierarchy were constructed in \cite{DJKM82a}, see also
\cite{N92}. A detailed discussion
of quasi-periodic solutions to the CKP hierarchy can be found in \cite{KZ20}.
The corresponding algebraic curves should admit a holomorphic involution with two
fixed points. 
Double-periodic in the complex plane (elliptic) solutions were studied in 
\cite{RZ20,Z19} for BKP and \cite{KZ20} for CKP. 

We also discuss the zero dispersion limit of the BKP and CKP hierarchies which appears to be
the same for both of them. In the dispersionless limit, the operator
$\p_x$ entering the pseudo-differential Lax operator
is replaced by a commuting variable $p$, the Lax operator becomes a commuting
function (a Laurent series) and the commutator is replaced by the Poisson bracket
$\{p,x\}=1$.

\section{The KP hierarchy}

Here we briefly recall the main notions related to the KP hierarchy.
The set of independent variables (``times'') is ${\bf t}=\{t_1, t_2, t_3, \ldots \}$.
It is convenient
to set $t_1 = x+\mbox{const}$, so that the vector fields $\p_{t_1}$ and $\p_x$ are 
identical: $\p_{t_1}=\p_x$. 
The main object is the $L$-operator which is a pseudo-differential operator of the form
\beq\label{kp1}
L=\p_x +u_1\p_x^{-1}+u_2 \p_x^{-2}+\ldots
\eeq
with no restrictions on the coefficient functions $u_i$.
The coefficient functions depend on $x$ and on all the times: 
$u_i=u_i(x, {\bf t})$. Together with
the Lax operator, it is convenient to introduce the wave operator (or dressing operator)
\beq\label{bkp1a}
W=1+\xi_1 \p_x^{-1}+\xi_2 \p_x^{-2}+\ldots
\eeq
such that
\beq\label{bkp1bb}
L=W\p_x W^{-1} 
\eeq
(the latter equality is interpreted as ``dressing'' of the operator $\p_x$ by $W$).
Clearly, there is a freedom in the definition of the wave operator: it can be multiplied
from the right by any pseudo-differential operator with constant coefficients.

The functions $u_i(x,0)$ are initial conditions for the time evolution 
$u_i(x,0)\to u_i(x, {\bf t})$
which is generated
by the Lax equations of the KP hierarchy:
\beq\label{kp3}
\p_{t_k}L=[B_k, \, L], \quad B_k = 
\Bigl (L^k\Bigr )_+, \quad k=1,2,3, \ldots ,
\eeq
where $(\ldots )_+$ means the differential part of a pseudo-differential operator
(i.e. terms with non-negative powers of $\p_x$). 
In particular,
$B_1=\p_x$ and $B_2=\p_x^2+2u_1$. 
Since $B_1=\p_x$, it follows from (\ref{kp3}) that the evolution in the time
$t_1$ is simply the shift of $x$, i.e. the solutions depend on $x+t_1$. 

An equivalent formulation of the hierarchy is through the zero curvature (Zakharov-Shabat)
equations
\beq\label{kp3a}
\p_{t_l}B_k-\p_{t_k}B_l+[B_k, B_l]=0.
\eeq
The equivalence of the Lax and Zakharov-Shabat formulations was proved in \cite{T92}.
The famous KP equation for $u_1$ is obtained from (\ref{kp3a}) at $k=2$, $l=3$. 

The Lax equations and the zero curvature equations 
are compatibility conditions of the auxiliary linear problems 
\beq\label{kp4}
\p_{t_k}\psi =B_k \psi , \quad L\psi =z\psi
\eeq
for the formal wave function 
\beq\label{kp4a}
\psi=\psi(x, {\bf t}, z)=We^{xz+\xi ({\bf t}, z)},
\eeq
where $W$ is the wave operator (\ref{bkp1a}), $z$ is the spectral parameter and
\beq\label{kp7a}
\xi ({\bf t}, z)=\sum_{k\geq 1}t_k z^k
\eeq
(it is implied that the operator $\p_x^{-1}$ acts to the exponential function as
$\p_x^{-1}e^{xz}=z^{-1}e^{xz}$). One can also introduce the adjoint (dual) wave function
\beq\label{kp4ab}
\psi^{\dag}=\psi^{\dag}(x, {\bf t}, z)=(W^{\dag})^{-1}e^{-xz-\xi ({\bf t}, z)},
\eeq
where $\dag$ means the formal adjoint 
defined by the rule $\Bigl (f(x)\circ \p_x^{n}\Bigr )^{\dag}=(-\p_x)^n \circ f(x)$.
It can be shown that the adjoint wave function satisfies the adjoint linear equations
\beq\label{kp4ad}
-\p_{t_k}\psi^{\dag} =B^{\dag}_k \psi^{\dag}.
\eeq

The tau-function $\tau^{\rm KP}(x, {\bf t})$ of the KP hierarchy
is consistently introduced
by the equations
\beq\label{ch1}
\psi (x, {\bf t}, z)=
e^{xz+\xi ({\bf t}, z)}
\frac{\tau^{\rm KP} (x, {\bf t}-[z^{-1}] )}{\tau^{\rm KP} (x, {\bf t})},
\eeq
\beq\label{ch1a}
\psi^{\dag}(x, {\bf t}, z)=
e^{-xz-\xi ({\bf t}, z)}
\frac{\tau^{\rm KP} (x, {\bf t}+[z^{-1}] )}{\tau^{\rm KP} (x, {\bf t})},
\eeq
where we have used the standard notation
$$
{\bf t}+j[z^{-1}]=
\Bigl \{ t_1+\frac{j}{z}, t_2+\frac{j}{2z^2}, t_3+\frac{j}{3z^3}, \ldots \Bigr \}, \quad
j\in \ZZ .
$$
The wave functions satisfy the bilinear equation \cite{JimboMiwa}
\beq\label{ch2}
\oint_{C_{\infty}}\psi (x, {\bf t}, z)
\psi^{\dag} (x, {\bf t}', z) \frac{dz}{2\pi i}=0
\eeq
for all ${\bf t}$, ${\bf t}'$. Here $C_{\infty}$ is a contour surrounding $\infty$
(a big circle of radius $R\to \infty$). 
Using (\ref{ch1}), (\ref{ch1a}), one can rewrite (\ref{ch2}) as the following bilinear
relation for the tau-function:
\beq\label{ch2a}
\oint_{C_{\infty}}e^{\xi ({\bf t}-{\bf t}', z)}\tau \Bigl (x, {\bf t}-[z^{-1}]\Bigr )
\tau \Bigl (x, {\bf t}'+[z^{-1}]\Bigr )\, \frac{dz}{2\pi i}=0.
\eeq
This is the generating equation for all differential equations of the KP hierarchy.
A direct consequence of the bilinear relation (\ref{ch2}) is the Hirota-Miwa equation
for the tau-function of the KP hierarchy 
\beq\label{tau6b}
\begin{array}{l}
(z_1-z_2)\tau^{\rm KP} \Bigr (x,{\bf t}-[z_1^{-1}]-[z_2^{-1}]\Bigl )
\tau^{\rm KP} \Bigr (x,{\bf t}-[z_3^{-1}]\Bigl )
\\ \\
\phantom{aaaaa}
+(z_2-z_3)\tau^{\rm KP} \Bigr (x,{\bf t}-[z_2^{-1}]-[z_3^{-1}]\Bigl )
\tau^{\rm KP} \Bigr (x,{\bf t}-[z_1^{-1}]\Bigl )
\\ \\
\phantom{aaaaaaaaaa}
+ (z_3-z_1)\tau^{\rm KP} \Bigr (x,{\bf t}-[z_1^{-1}]-[z_3^{-1}]\Bigl )
\tau^{\rm KP} \Bigr (x,{\bf t}-[z_2^{-1}]\Bigl )=0.
\end{array}
\eeq
It is a generating equation for the differential equations of the hierarchy. 
In the limit $z_3\to \infty$ it becomes the equation
\beq\label{ch3}
\begin{array}{l}
\displaystyle{
\p_{x}\log \frac{\tau^{\rm KP}\Bigl (x, {\bf t}+
[z_1^{-1}]-[z_2^{-1}]\Bigr )}{\tau^{\rm KP}(x, {\bf t})}}
\\ \\
\phantom{aaaaaaaaaaaaa}\displaystyle{=
(z_2-z_1)\left (\frac{\tau^{\rm KP}\Bigl (x, {\bf t}+
[z_1^{-1}]\Bigr )\tau^{\rm KP}\Bigl (x, {\bf t}-
[z_2^{-1}]\Bigr )}{\tau^{\rm KP}(x, {\bf t})
\tau^{\rm KP}\Bigl (x, {\bf t}+
[z_1^{-1}]-[z_2^{-1}]\Bigr )}-1\right )}.
\end{array}
\eeq

The tau-function $\tilde \tau (x, {\bf t})=e^{\ell (x,{\bf t})}\tau (x, {\bf t})$, where
$\displaystyle{\ell (x, {\bf t})=\gamma_0 +\gamma_1 x +\sum_{k\geq 1}\gamma_k t_k}$ is a linear
function of the times, satisfies the same bilinear equations. We say that the tau-functions
which differ by a factor of the form $e^{\ell (x,{\bf t})}$ are equivalent.

\section{The BKP hierarchy}

Here we present the main formulas 
related to the BKP hierarchy with some details. The main reference is \cite{DJKM82}, see also
\cite{JimboMiwa,LW99,Tu07,DM-H09}.

\subsection{The Lax operator and the wave operator}

The set of independent variables (``times'') is ${\bf t}_{\rm o}=\{t_1, t_3, t_5, \ldots \}$. 
They are indexed by positive odd numbers. As in the KP hierarchy, we
set $t_1 = x+\mbox{const}$, so that the vector fields $\p_{t_1}$ and $\p_x$ are 
identical: $\p_{t_1}=\p_x$. 
The main object is the $L$-operator which is a pseudo-differential operator of the form
\beq\label{bkp1}
L=\p_x +u_1\p_x^{-1}+u_2 \p_x^{-2}+\ldots
\eeq
with the constraint
\beq\label{bkp2}
L^{\dag}=-\p_x L\p_x^{-1}.
\eeq
Unlike in the case of a reduction, when only a finite number of the coefficient functions
$u_i$ remain independent, the constraint (\ref{bkp2}) implies that there are 
still infinitely many independent coefficients functions. 
As we shall see soon,
the constraint (\ref{bkp2}) 
is invariant under the ``odd'' flows $t_1, t_3, t_5, \ldots$ of the KP
hierarchy, so the BKP hierarchy can be regarded as a sub-hierarchy 
(a restriction) of the KP one with the 
``even'' times frozen.

It is instructive to reformulate the constraint (\ref{bkp2}) in terms of the
wave operator $W$ (\ref{bkp1a}) such that $L=W\p_x W^{-1}$. 
The constraint (\ref{bkp2}) implies that $W^{\dag}\p_x W$ commutes with $\p_x$, i.e., 
it is a pseudo-differential operator with constant coefficients. 
The freedom in the definition of the wave operator can be fixed by demanding that
$W^{\dag}\p_x W=\p_x$, i.e.
\beq\label{bkp1b}
W^{\dag}=\p_x W^{-1}\p_x^{-1}.
\eeq

The Lax equations of the hierarchy are the same as (\ref{kp3}) but with odd indices:
\beq\label{bkp3}
\p_{t_k}L=[B_k, \, L], \quad B_k = 
\Bigl (L^k\Bigr )_+, \quad k=1,3,5, \ldots .
\eeq
The constraint (\ref{bkp2}) is equivalent to the condition that
the differential operators $B_k$ satisfy $B_k\cdot 1=0$ (for odd $k$), i.e.,
that they have the form
$$
B_k=\p_x^k +\sum_{j=1}^{k-2}b_{k,j}\p_x^{j}
$$
with $b_{k,0}=0$. 
Indeed, if (\ref{bkp2}) is satisfied, then $L^n \p_x^{-1}=-\p_x^{-1}L^{\dag n}$ for odd $n$.
On the other hand, $(L^n \p_x^{-1})^{\dag}=-\p_x^{-1}L^{\dag n}=L^n \p_x^{-1}$ which implies that
the coefficient in front of $\p_x^0$ in $L^n$ vanishes: $b_{n,0}=0$ for odd $n$.
Conversely, assuming that $b_{n,0}=0$ for all odd $n$, we shall prove that 
$R=\p_x^{-1}L^{\dag}+L\p_x^{-1}=0$. Obviously, $R$ is of the general form
$$
R=a\p_x^{-m}+\mbox{lower order terms}
$$
and the identity $R^{\dag}=-R$ implies that if $a$ is not identically zero, then $m$ is odd. 
Then we have (see \cite{DJKM83}):
$$
L^m \p_x^{-1}=(R\p_x -\p_x^{-1}L^{\dag}\p_x )^m \p_x^{-1}
$$
$$
=-(\p_x^{-1}L^{\dag}\p_x)^m \p_x^{-1}+mR (\p_x^{-1}L^{\dag}\p_x)^{m-1} +
\mbox{an operator of order less than $-1$}
$$
$$
=(L^n \p_x^{-1})^{\dag}+ma \p_x^{-1} +\mbox{lower order terms}
$$
which contradicts the assumption that $a\neq 0$.

Note that the constraint (\ref{bkp2}) implies $(L^{\dag})^k_{+}=-(\p_x L^k \p_x^{-1})_{+}\! =\! 
-(L^k)_{+}\! -\! ((\p_x L^k)\p_x^{-1})_{+}$ which can be rewritten as
\beq\label{bkp101}
B_k^{\dag}=-\p_x B_k \p_x^{-1}, \qquad \mbox{$k$ odd}
\eeq
(taking into account that $b_{k,0}=0$).
Using this relation, it is straightforward to check, using the Lax equations,
that the constraint (\ref{bkp2})
is indeed invariant under odd flows of the KP hierarchy:
\beq\label{bkp102}
\p_{t_k}\Bigl (L^{\dag}+\p_x L \p_x^{-1}\Bigr )=0, \qquad \mbox{$k$ odd}.
\eeq
Therefore, the BKP hierarchy is well-defined as a subhierarchy of the KP hierarchy.

The first three differential operators $B_k$ are as follows:
\beq\label{bkp5}
\begin{array}{l}
B_1=\p_x,
\\ \\
B_3 =\p_x^3 +6u\p_x, \quad u=\frac{1}{2}\, u_1,
\\ \\
B_5 =\p_x^5 +10u\p^3_x +10u'\p_x^2+v\p_x.
\end{array}
\eeq
An equivalent formulation of the hierarchy is through the zero curvature 
equations
\beq\label{bkp3a}
\p_{t_l}B_k-\p_{t_k}B_l+[B_k, B_l]=0, \quad \mbox{$k,l$ odd}.
\eeq
The first equation of the BKP hierarchy follows from the zero curvature
equation $\p_{t_3}B_5-\p_{t_5}B_3+[B_5, B_3]=0$. The calculations yield the following
system of equations for the unknown functions $u,v$:
\beq\label{bkp0}
\left \{
\begin{array}{l}
3v' =10 u_{t_3}+20 u^{'''} +120 uu'
\\ \\
v_{t_3}-6u_{t_5}=v^{'''}-6u^{'''''}-60 uu^{'''}-60 u'u'' +6uv'-6vu',
\end{array}
\right.
\eeq
Note that the variable $v$ can be excluded by passing to the unknown function $U$
such that $U'=u$. 

\subsection{The wave function and the tau-function}

\label{section:wavetau}

The Lax equations and the zero curvature equations 
are compatibility conditions of the auxiliary linear problems (\ref{kp4})
for the formal wave function 
\beq\label{bkp4a}
\psi=\psi(x, {\bf t}_{\rm o}, z)=We^{xz+\xi ({\bf t}_{\rm o}, z)},
\eeq
where 
\beq\label{bkp7a}
\xi ({\bf t}_{\rm o}, z)=\sum_{k\geq 1, \, {\rm odd}}t_k z^k.
\eeq
As it follows from (\ref{bkp1a}), 
the wave function $\psi =\psi(x,{\bf t}_{\rm o}, z)$ has the following expansion as $z\to \infty$:
\beq\label{bkp5b}
\psi(x, {\bf t}_{\rm o}, z)=e^{xz+\xi ({\bf t}_{\rm o}, z)} 
\Bigl (1+\sum_{k\geq 1}\xi_k z^{-k}\Bigr ).
\eeq
As is proved in \cite{DJKM81}, the wave function  satisfies the
bilinear relation
\beq\label{bkp5a}
\oint_{C_{\infty}}\! \psi (x, {\bf t}_{\rm o}, z)\psi (x, {\bf t}_{\rm o}', -z)\frac{dz}{2\pi iz}=1
\eeq
valid for all ${\bf t}_{\rm o}, {\bf t}_{\rm o}'$.
For completeness, we give a sketch of proof here.
By virtue of the differential equations (\ref{bkp3}), the bilinear relation is equivalent
to vanishing of
$$
b_m=\p_{x'}^m \oint_{C_{\infty}}\! \psi (x, {\bf t}_{\rm o}, z)\psi (x', {\bf t}_{\rm o}, -z)
\frac{dz}{2\pi iz} \Biggr |_{x'=x}\quad \mbox{for all $m\geq 1$}
$$
with the additional condition that
$$
b_0=\oint_{C_{\infty}}\! \psi (x, {\bf t}_{\rm o}, z)\psi (x, {\bf t}_{\rm o}, -z)
\frac{dz}{2\pi iz} =1.
$$
We have:
$$
b_m=\oint_{C_{\infty}}\Bigl (\sum_{k\geq 0}\xi_k(x)z^{-k}\Bigr )\p_{x'}^m
\Bigl (\sum_{l\geq 0}\xi_l(x')(-z)^{-l}\Bigr )e^{(x-x')z} \frac{dz}{2\pi iz} \Biggr |_{x'=x}
$$
$$
=\oint_{C_{\infty}}\Bigl (\sum_{k\geq 0}\xi_kz^{-k}\Bigr )(\p_x -z)^m
\Bigl (\sum_{l\geq 0}\xi_l(-z)^{-l}\Bigr )\frac{dz}{2\pi iz} 
$$
$$
=\sum_{j+k+l=m}(-1)^{m+j+l}\left (\! \begin{array}{c}m\\ j \end{array} \! \right )
\xi_k \p_x^j \xi_l.
$$
But the last expression is the coefficient of $(-1)^m \p_x^{-m-1}$ in the operator
$W\p_x^{-1}W^{\dag}$:
$$
W\p_x^{-1}W^{\dag}=\p_x^{-1}+\sum_{m\geq 1}(-1)^m b_m \p_x^{-m-1}.
$$
Since $W\p_x^{-1}W^{\dag}=\p_x^{-1}$, we conclude that $b_m=0$ for all $m\geq 1$
and $b_0=1$. 

The tau-function $\tau = \tau (x, {\bf t}_{\rm o})$ 
of the BKP hierarchy is consistently introduced by the formula 
\beq\label{bkp6}
\psi  = e^{xz+\xi ({\bf t}_{\rm o}, z)}\,
\frac{\tau (x, {\bf t}_{\rm o}-2[z^{-1}]_{\rm o})}{\tau (x, {\bf t}_{\rm o})},
\eeq
where
\beq\label{bkp8a}
{\bf t}_{\rm o}+k[z^{-1}]_{\rm o} \equiv \Bigl \{ t_1 +\frac{k}{z},  t_3 +\frac{k}{3z^3}, 
t_5 +\frac{k}{5z^5}, \, \ldots \Bigr \}, \quad k\in \ZZ .
\eeq
The proof of the existence of the tau-function is based on the bilinear relation. 
Let us represent the wave function
in the form
$$
\psi (x, {\bf t}_{\rm o}, z)=e^{xz+\xi ({\bf t}_{\rm o}, z)}w(x,{\bf t}_{\rm o}, z)
$$
and set ${\bf t}_{\rm o}'={\bf t}_{\rm o}-2[a^{-1}]_{\rm o}$ in the bilinear relation. We have
$\displaystyle{e^{\xi ({\bf t}_{\rm o}-{\bf t}_{\rm o}', z)}=
\frac{a+z}{a-z}}$ and the residue calculus
yields
\beq\label{bder1}
w({\bf t}_{\rm o}, a) w({\bf t}_{\rm o}-2[a^{-1}]_{\rm o}, -a)=1,
\eeq
where
we do not indicate the dependence on $x$ for brevity. 
Next, we set
${\bf t}_{\rm o}'={\bf t}_{\rm o}-2[a^{-1}]_{\rm o}-2[b^{-1}]_{\rm o}$ in the
bilinear relation, so that $\displaystyle{e^{\xi ({\bf t}_{\rm o}-{\bf t}_{\rm o}', z)}=
\frac{(a+z)(b+z)}{(a-z)(b-z)}}$.
In this case the residue calculus yields
\beq\label{bder7}
w({\bf t}_{\rm o}, a)w({\bf t}_{\rm o}-2[a^{-1}]_{\rm o}-2[b^{-1}]_{\rm o}, -a)=
w({\bf t}_{\rm o}, b)w({\bf t}_{\rm o}-2[a^{-1}]_{\rm o}-2[b^{-1}]_{\rm o}, -b).
\eeq
With the help of (\ref{bder1}) this latter relation can be rewritten as
\beq\label{bder8}
\frac{w({\bf t}_{\rm o}, a)
w({\bf t}_{\rm o}-2[a^{-1}]_{\rm o}, b)}{w({\bf t}_{\rm o}, b)
w({\bf t}_{\rm o}-2[b^{-1}]_{\rm o}, a)}=1.
\eeq

Now we are going to show that (\ref{bder8}) 
implies that there exists a function $\tau ({\bf t}_{\rm o})$ such that
\beq\label{bder9}
w({\bf t}_{\rm o}, z)=\frac{\tau ({\bf t}_{\rm o}-2[z^{-1}]_{\rm o})}{\tau ({\bf t}_{\rm o})}.
\eeq
To see this, let us represent (\ref{bder9}) in an equivalent form. 
Taking logarithm and $z$-derivative, we have from (\ref{bder9})
$$
\p_z \log w =2\! \sum_{m\geq 1, \,\, {\rm odd}}\! z^{-m-1}\p_{t_m}\log
\tau ({\bf t}_{\rm o}-2[z^{-1}]_{\rm o}),
$$
or, substituting $\tau ({\bf t}_{\rm o}-2[z^{-1}]_{\rm o})$ expressed through
$w({\bf t}_{\rm o}, z)$ and $\tau ({\bf t}_{\rm o})$ from (\ref{bder9}) in the right hand side,
\beq\label{bder10}
\p_z \log w = 2\p_{{\bf t}_{\rm o}} (z)\log w +
2\p_{{\bf t}_{\rm o}} (z)\log \tau ,
\eeq
where $\p_{{\bf t}_{\rm o}} (z)$ is the differential operator
$$
\p_{{\bf t}_{\rm o}} (z)=\sum_{j \,\, {\rm odd}}z^{-j-1}\p_{t_j}.
$$
In fact (\ref{bder10}) is equivalent to (\ref{bder9}). Indeed, writing 
(\ref{bder10}) as $(\p_z-2\p_{{\bf t}_{\rm o}} (z))\log (w\tau )=0$, we conclude that
$w\tau =\rho$ is a function of ${\bf t}_{\rm o}-2[z^{-1}]_{\rm o}$, and the normalization
condition $w({\bf t}_{\rm o}, \infty )=1$ implies that $\rho =\tau$, so
we arrive at (\ref{bder9}). 

Equation (\ref{bder9}) means that
$$
Y_n :=\res_{z=\infty} \Bigl [ z^n (\p_z -2\p_{{\bf t}_{\rm o}} (z))\log w \Bigr ]=
2\frac{\p \log \tau}{\p t_n},
$$
where the residue is defined as $\res_{z=\infty}(z^{n-1}) =\delta_{n0}$. 
Therefore, the existence of the tau-function will be proved if we prove that
$\p_{t_n}Y_m({\bf t}_{\rm o})=\p_{t_m}Y_n({\bf t}_{\rm o})$. Changing $a\to z$,
$b\to \zeta$ in (\ref{bder8}), and applying the operator 
$\p_z -2\p_{{\bf t}_{\rm o}} (z)$ to logarithm of this equality, we rewrite it as
$$
(\p_z \! -\! 2\p_{{\bf t}_{\rm o}} (z))\log w({\bf t}_{\rm o}, z)-
(\p_z \! -\! 2\p_{{\bf t}_{\rm o}} (z))\log w({\bf t}_{\rm o}\! -\! 2[\zeta ^{-1}]_{\rm o}, z)=
-2\p_{{\bf t}_{\rm o}} (z) \log w_0({\bf t}_{\rm o}, \zeta ),
$$
or
\beq\label{bder11}
Y_n ({\bf t}_{\rm o})-Y_n ({\bf t}_{\rm o}-2[\zeta^{-1}]_{\rm o})=-2
\p_{t_n}\log w({\bf t}_{\rm o}, \zeta ).
\eeq 
Therefore, denoting $F_{mn}=\p_{t_m}Y_n-\p_{t_n}Y_m$, we see from (\ref{bder11})
that
\beq\label{bder12}
F_{mn}({\bf t}_{\rm o})=F_{mn}({\bf t}_{\rm o}-2[\zeta^{-1}]_{\rm o}).
\eeq
This equality is valid identically in $\zeta$.
Expanding its right hand side in a power series,
$$
\begin{array}{c}
F_{mn}({\bf t}_{\rm o}\! -\! 2[\zeta^{-1}]_{\rm o})=F_{mn}({\bf t}_{\rm o})
\! -\! 2\zeta^{-1}\p_{t_1}F_{mn}({\bf t}_{\rm o})\! -\!  \frac{2}{3}\, \zeta^{-3}(\p_{t_3}
F_{mn}({\bf t}_{\rm o})\! +\! 2 \p_{t_1}^3F_{mn}({\bf t}_{\rm o}))+\ldots ,
\end{array}
$$
we conclude from compating of the $\zeta^{-1}$-terms 
that $F_{mn}$ does not depend on $t_1$. From 
the $\zeta^{-3}$-terms we see that it does not depend on $t_3$ and so on. In this way
we can conclude that it
does not depend on $t_k$ for all (odd) $k$, i.e.
$F_{mn}=2a_{mn}$, where $a_{mn}$ are some constants such that $a_{mn}=-a_{nm}$. 
Therefore, we can write
$$
Y_n =\sum_m a_{mn}t_m +\p_{t_n}h,
$$
with some function $h=h({\bf t}_{\rm o})$. Then from (\ref{bder11}) we have
$$
-2\p_{t_n}\log w({\bf t}_{\rm o}, z)=\p_{t_n}(h({\bf t}_{\rm o})-
h({\bf t}_{\rm o}-2[z^{-1}]_{\rm o}))+2\! \sum_{m\,\, {\rm odd}}\frac{a_{mn}}{m}\, z^{-m},
$$
or, after integration,
$$
\log w({\bf t}_{\rm o}, z)=\frac{1}{2}\, h({\bf t}_{\rm o}\! -\! 
2[z^{-1}]_{\rm o}) -\frac{1}{2}\, h({\bf t}_{\rm o})-
\sum_{m\,\, {\rm odd}}\frac{a_{mn}}{m}\, z^{-m}t_n
+\varphi (z),
$$
where $\varphi (z)$ is a function of $z$ only. Substituting this into 
(\ref{bder8}), we conclude that $a_{mn}=0$, and so
$\p_{t_m}Y_n =\p_{t_n}Y_m$. 

Let us show how to obtain (\ref{bkp6}) up to a common $x$-independent factor 
in a very easy way.
Apply $\p_{t_1}$ to
(\ref{bkp5a}) and set ${\bf t}_{\rm o}'={\bf t}_{\rm o}-2[a^{-1}]_{\rm o}$. 
The residue calculus yields
\beq\label{bder3}
\begin{array}{c}
2a\Bigl (w({\bf t}_{\rm o}, a) w({\bf t}_{\rm o}\! -\! 2[a^{-1}]_{\rm o}, -a)-1\Bigr )+2
w'({\bf t}_{\rm o}, a) w({\bf t}_{\rm o}-2[a^{-1}]_{\rm o}, -a)
\\ \\
+\xi_1({\bf t}_{\rm o}\! -\! 2[a^{-1}]_{\rm o})-\xi_1({\bf t}_{\rm o})=0.
\end{array}
\eeq
Using (\ref{bder1}), we conclude from (\ref{bder3}) that
\beq\label{bder6}
\p_x \log w({\bf t}_{\rm o}, a)
=\frac{1}{2}\Bigl (\xi_1({\bf t}_{\rm o})-\xi_1({\bf t}_{\rm o}-2[a^{-1}]_{\rm o})\Bigr ).
\eeq
Now, setting $\xi_1(x, {\bf t}_{\rm o})=-2\p_x\log 
\tau (x, {\bf t}_{\rm o})$ and integrating, we arrive at (\ref{bkp6}) up
to a common $x$-independent factor.

The function $u$ in (\ref{bkp5}) can be also expressed
through the tau-function with the help of the following argument.
It is a matter of direct verification that the result of the action of the operator
$\p^3_x +6u\p_x -\p_{t_3}$ to the wave function $\psi$ of the form (\ref{bkp5b}) is 
$O(z^{-1})$ as $z\to \infty$, i.e., 
\beq\label{bkp5e}
(\p^3_x +6u\p_x -\p_{t_3})\psi =O(z^{-1})e^{xz+\xi ({\bf t}_{\rm o}, z)}
\eeq
if the conditions
\beq\label{bkp5d}
u=-\frac{1}{2}\, \xi_1', \quad \xi_1\xi_1' -\xi_1'' -\xi_2'=0
\eeq
hold true (actually we know that $(\p^3_x +6u\p_x -\p_{t_3})\psi =0$ but here
we only need the weaker condition (\ref{bkp5e})). Since from (\ref{bkp6}) it follows that
$$\xi_1=-2\p_x\log \tau , \quad \xi_2=2(\p_x \log \tau )^2 +2\p_x^2\log \tau ,$$ we have
\beq\label{bkp10}
u=\p_x^2\log \tau 
\eeq
and the second equality in (\ref{bkp5d}) holds identically. 

The change of dependent variables from $u,v$ to the tau-function 
as in (\ref{bkp10}) and
\beq\label{bkp01}
v=\frac{10}{3}\, \p_{t_3}\p_x \log \tau +
\frac{20}{3}\, \p_x^4 \log \tau +20 (\p_x^2 \log \tau )^2
\eeq
makes the first of the equations (\ref{bkp0}) trivial 
and the other one turns into the bilinear form \cite{DJKM82}
\beq\label{bkp02}
\Bigl (D_1^6 -5D_1^3D_3-5D_3^2+9D_1D_5 \Bigr ) \tau \cdot \tau =0,
\eeq
where $D_i$ are the Hirota operators defined by the rule
$$
\begin{array}{l}
P(D_1, D_3, D_5, \ldots )\tau \cdot \tau 
\\ \\
\phantom{aaaaaaa}=
P(\p_{y_1}, \p_{y_3}, \p_{y_5}, \ldots )
\tau (x, t_1+y_1, t_3+y_3, \ldots )\tau (x, t_1-y_1, t_3-y_3, \ldots )\Bigr |_{y_i=0}
\end{array}
$$
for any polynomial $P(D_1, D_3, D_5, \ldots )$.

As it follows from (\ref{bkp6}), 
the BKP hierarchy is equivalent to the following relation for the 
tau-function:
\beq\label{bkp9b}
\oint_{C_{\infty}} \!\! e^{\xi ({\bf t}_{\rm o}-{\bf t}_{\rm o}', z)}
\tau \Bigl (x,{\bf t}_{\rm o}-2[z^{-1}]_{\rm o}\Bigr )\tau 
\Bigl (x, {\bf t}_{\rm o}'+2[z^{-1}]_{\rm o}\Bigr )
\, \frac{dz}{2\pi iz} =\tau (x, {\bf t}_{\rm o})\tau (x, {\bf t}_{\rm o}')
\eeq
valid for all ${\bf t}_{\rm o}, {\bf t}_{\rm o}'$.  Set 
${\bf t'}_{\rm o}={\bf t}_{\rm o}-2[a^{-1}]_{\rm o}-
2[b^{-1}]_{\rm o}-2[c^{-1}]_{\rm o}$, then
$$
e^{\xi ({\bf t}_{\rm o}-{\bf t}_{\rm o}', z)}=\frac{(a+z)(b+z)(c+z)}{(a-z)(b-z)(c-z)}
$$
and the residue calculus in (\ref{bkp9b}) gives the following equation:
\beq\label{bkp11}
\begin{array}{c}
(a+b)(a+c)(b-c)\tau^{[a]}\tau^{[bc]}
+ (b+a)(b+c)(c-a)\tau^{[b]}\tau^{[ac]}
\\ \\
+ (c+a)(c+b)(a-b)\tau^{[c]}\tau^{[ab]}
+(a-b)(b-c)(c-a)\tau \tau^{[abc]}=0,
\end{array}
\eeq
where
$
\tau^{[a]}=\tau (x, {\bf t_{\rm o}}+2[a^{-1}]_{\rm o})$,
$\tau^{[ab]}=\tau (x, {\bf t}_{\rm o}+2[a^{-1}]_{\rm o}+2[b^{-1}]_{\rm o})$,
and so on.
Equation (\ref{bkp11}) should be valid for all $a,b,c$. 
Taking the limit $c\to \infty$, we get the equation
\beq\label{bkp12}
\tau \tau^{[ab]}\left (1+\frac{1}{a+b}\, \p_{t_1}\log \frac{\tau}{\tau^{[ab]}}\right )
=\tau^{[a]}\tau^{[b]}
\left (1+\frac{1}{a-b}\, \p_{t_1}\log \frac{\tau^{[b]}}{\tau^{[a]}}\right )
\eeq
This is the equation for the tau-function of the BKP hierarchy. It should hold for all
$a,b$. The differential equations of the hierarchy are obtained by expanding it in 
inverse powers of $a,b$.

\subsection{The BKP hierarchy and the KP hierarchy}

Here we compare the spaces of solutions to the BKP and KP hierarchies by
an explicit embedding of the former into the latter on the level of 
tau-functions.

We begin with the adjoint wave function:
\beq\label{comp1}
\psi^{\dag}(x, {\bf t}_{\rm o}, z)=(W^{\dag})^{-1}e^{-xz-\xi ({\bf t}_{\rm o}, z)}
=\p_x W\p_x^{-1}e^{-xz-\xi ({\bf t}_{\rm o}, z)}=-z^{-1}\p_x \psi (x, {\bf t}_{\rm o}, -z),
\eeq
so that the bilinear relation (\ref{ch2}) reads
\beq\label{comp2}
\oint_{C_{\infty}}\! \psi (x, {\bf t}_{\rm o}, z)\p_x\psi (x, {\bf t}_{\rm o}', -z)
\frac{dz}{2\pi iz}=0
\eeq
which is a consequence of (\ref{bkp5a}). Using (\ref{ch1}), (\ref{ch1a}), one can 
write (\ref{comp1}) in terms of the KP tau-function:
$$
e^{-xz}\frac{\tau^{\rm KP} (x,{\bf \dot t}+[z^{-1}])}{\tau^{\rm KP} (x,{\bf \dot t})}=
-z^{-1}\p_x \left (\frac{e^{-xz}
\tau^{\rm KP} (x,{\bf \dot t}-[-z^{-1}])}{\tau^{\rm KP} (x,{\bf \dot t})}\right )
$$
or
\beq\label{comp3}
\p_x \log \frac{\tau^{\rm KP} (x,{\bf \dot t}-[z^{-1}])}{\tau^{\rm KP} (x,{\bf \dot t})}
=-z\left (1-\frac{\tau^{\rm KP}(x,{\bf \dot t}+[-z^{-1}])}{\tau^{\rm KP}(x,{\bf \dot t}-[z^{-1}])}
\right ),
\eeq
where we use the short-hand notation ${\bf \dot t}=\{t_1, 0, t_3, 0, \ldots \}$.
Shifting the times ${\bf t}_{\rm o}$, we can rewrite this as
$$
\p_x \log \frac{\tau^{\rm KP}(t_1, -\frac{1}{2}\, z^{-2}, 
t_3, -\frac{1}{4}\, z^{-4},\ldots )}{\tau^{\rm KP}(t_1+z^{-1}, 0, t_3+\frac{1}{3}\, z^{-3}, 0,
\ldots )}=-z +z 
\frac{\tau^{\rm KP}(t_1, \frac{1}{2}\, z^{-2}, 
t_3, \frac{1}{4}\, z^{-4},\ldots )}{\tau^{\rm KP}(t_1, -\frac{1}{2}\, z^{-2}, 
t_3, -\frac{1}{4}\, z^{-4},\ldots )},
$$
or, subtracting these equalities with $z$ and $-z$,
\beq\label{comp4}
\p_x \log \frac{\tau^{\rm KP}(x,t_1-z^{-1}, 0, t_3-\frac{1}{3}\, z^{-3}, 0,
\ldots )}{\tau^{\rm KP}(x,t_1+z^{-1}, 0, t_3+\frac{1}{3}\, z^{-3}, 0,
\ldots )}=-2z +2z 
\frac{\tau^{\rm KP}(x,t_1, \frac{1}{2}\, z^{-2}, 
t_3, \frac{1}{4}\, z^{-4},\ldots )}{\tau^{\rm KP}(x,t_1, -\frac{1}{2}\, z^{-2}, 
t_3, -\frac{1}{4}\, z^{-4},\ldots )}.
\eeq
Comparing this with the KP hierarchy in the form (\ref{ch3}) at $z_2=-z_1=z$,
we conclude that
\beq\label{comp5}
\Bigl (\tau^{\rm KP}(x,{\bf \dot t}\! -\! [z^{-1}])\Bigr )^2=\tau^{\rm KP}(x,{\bf \dot t})
\tau^{\rm KP}(x,{\bf \dot t}-2[z^{-1}]_{\rm o})
\eeq
(in the second tau-function in the right hand side, the even times are equal to 0).
This is the constraint which distinguishes solutions to the BKP hierarchy among all
solutions to the KP hierarchy.

Another way to come to (\ref{comp5}) is to notice 
that we have two different expressions for the wave function $\psi$
(one in terms of the KP tau-function $\tau^{\rm KP}$ and the other in terms of the BKP 
tau-function $\tau$)
from which it follows that
$$
\frac{\tau^{\rm KP} (x,{\bf \dot t}-[z^{-1}])}{\tau^{\rm KP} (x,{\bf \dot t})}=
\frac{\tau  (x,{\bf t}_{\rm o}-2[z^{-1}]_{\rm o})}{\tau (x,{\bf t}_{\rm o})}
$$
or, after a shift of the times ${\bf t}_{\rm o}$,
\beq\label{comp6}
\log \frac{\tau^{\rm KP}(x,t_1, -\frac{1}{2}\, z^{-2}, 
t_3, -\frac{1}{4}\, z^{-4}, \ldots )}{\tau^{\rm KP}(x,t_1+z^{-1}, 0, t_3+\frac{1}{3}\, z^{-3}, 0,
\ldots )}=
\log \frac{\tau  (x,{\bf t}_{\rm o}-[z^{-1}]_{\rm o})}{\tau (x,{\bf t}_{\rm o}-[-z^{-1}]_{\rm o})}.
\eeq
The right hand side is and odd function of $z$, therefore, we have
$$
\log \frac{\tau^{\rm KP}(x,t_1, -\frac{1}{2}\, z^{-2}, 
t_3, -\frac{1}{4}\, z^{-4}, \ldots )}{\tau^{\rm KP}(x,t_1+z^{-1}, 0, t_3+\frac{1}{3}\, z^{-3}, 0,
\ldots )}+\log \frac{\tau^{\rm KP}(x,t_1, -\frac{1}{2}\, z^{-2}, 
t_3, -\frac{1}{4}\, z^{-4}, \ldots )}{\tau^{\rm KP}(x,t_1-z^{-1}, 0, t_3-\frac{1}{3}\, z^{-3}, 0,
\ldots )}=0
$$
which is (\ref{comp5}). 

The constraint (\ref{comp5}) should be valid for all values of $t_1, t_3, \ldots $ and $z$.
Expanding it in powers of $z$, one can represent it as an infinite number of differential
constraints the first of which is
\beq\label{comp7}
(\p_{t_2}+\p_{t_1}^2)\log \tau^{\rm KP}\Bigr |_{t_{2k}=0}=0, \quad k\geq 1.
\eeq
This constraint was mentioned in \cite{N92}.

Let us represent (\ref{comp5}) in the form
\beq\label{comp8}
\frac{\tau^{\rm KP}(x,{\bf \dot t}-2[z^{-1}]_{\rm o})}{\tau^{\rm KP}
(x,{\bf \dot t})}=\left (\frac{\tau^{\rm KP}(x,{\bf \dot t}-[z^{-1}])}{\tau^{\rm KP}
(x,{\bf \dot t})}\right )^2=
\frac{\tau^2 (x, {\bf t}_{\rm o}-2[z^{-1}]_{\rm o})}{\tau^2(x, {\bf t}_{\rm o})}.
\eeq
It follows from here that
\beq\label{comp9}
\tau (x, {\bf t}_{\rm o})=C\sqrt{\vphantom{A^A}\tau^{\rm KP} (x, {\bf \dot t})}
\eeq
and that $\tau^{\rm KP} (x, {\bf \dot t})$ is a full square, i.e.,
$\sqrt{\vphantom{A^A}\tau^{\rm KP} (x, {\bf \dot t})}$ does not have square root singularities
in all the times.

\subsection{Examples: soliton solutions}

$N$-soliton solutions of the BKP hierarchy are obtained by imposing certain constraints on
the parameters of $2N$-soliton solutions to the KP hierarchy. The tau-function of the BKP
hierarchy is related to the KP tau-function as
$\tau =\sqrt{\vphantom{B^{a^a}}\tau^{\rm KP}}$, with ``even'' times $t_{2k}$ put equal to zero
and it is implied that the parameters of the KP tau-function $\tau^{\rm KP}$ 
are chosen in a special way. With this choice, $\tau^{\rm KP}$ is a full square, i.e.,
$\tau$ does not have square root singularities.

\subsubsection{One-soliton solution}

The tau-function for one BKP soliton is the square root 
of a specialization of 2-soliton tau-function 
of the KP hierarchy:
\beq\label{be2}
\tau^{\rm KP}\Bigr |_{t_{2k}=0}= 1+\alpha (p-q)w+
\frac{\alpha^2}{4}\, (p-q)^2 \, w^2 = \Bigl (1+\frac{\alpha}{2}\, (p-q)w\Bigr )^2,
\eeq
where 
\beq\label{be5}
w=e^{(p+q)x+ \xi ({\bf t}_{\rm o}, p)+\xi ({\bf t}_{\rm o}, q)}, \quad 
\mbox{$\xi ({\bf t}_{\rm o}, z)$ is given by (\ref{bkp7a})}
\eeq
and
$\alpha , p, q$ are arbitrary parameters. Therefore, the tau-function of the BKP 
hierarchy is
\beq\label{be2a}
\tau =1+\frac{\alpha}{2}\, (p-q)w.
\eeq 
It is an entire function of $x$.

Note that the extension of the tau-function $\tau^{\rm KP}$ to the modified KP hierarchy
(mKP) reads
\beq\label{e6}
\tau_n^{\rm mKP}= 1+\alpha \Bigl (-q(-p/q)^n +p(-q/p)^n \Bigr ) w+
\frac{\alpha^2}{4}\, (p-q)^2  w^2,
\eeq
where $n$ is the integer-valued ``zeroth time''
(clearly,
$\tau^{\rm KP}=\tau_0^{\rm mKP}=\tau_1^{\rm mKP}$). Then we see that the parameters of the
soliton solutions are such that $\tau_{1-n}^{\rm mKP}=\tau_n^{\rm mKP}$, which is the
constraint necessary for the BKP hierarchy \cite{JimboMiwa,UT84}.

The bilinear identity
(\ref{bkp5a}), which in the present case has the explicit form
$$
\oint_{C_{\infty}} \!\! e^{\xi ({\bf t}_{\rm o}, z)-\xi ({\bf t}_{\rm o}', z)}
\left (\! 1+\frac{\alpha}{2}\, \frac{(z-p)(z-q)}{(z+p)(z+q)}\, (p-q)w\right )\!
\left (\! 1+\frac{\alpha}{2}\, \frac{(z+p)(z+q)}{(z-p)(z-q)}\, (p-q)w'\right )
\frac{dz}{2\pi i z}
$$
$$
=\left (1+\frac{\alpha}{2}\, (p-q)w\right )\left (1+\frac{\alpha}{2}\, (p-q)w'\right )
$$
can be checked directly by the residue calculus. 
(Here $w'=e^{(p+q)x+ \xi ({\bf t}_{\rm o}', p)+\xi ({\bf t}_{\rm o}', q)}$.)

\subsubsection{Multi-soliton solutions}

The general KP tau-function for $2N$-soliton solution has $6N$ arbitrary parameters
$\alpha_i$, $p_i$, $q_i$ ($i=1, \ldots , 2N$) and is given by
\beq\label{ms1}
\begin{array}{c}
\displaystyle{
\tau^{\rm KP} \left [ \begin{array}{c}\alpha_1 \\ p_1, q_1\end{array};
\begin{array}{c}\alpha_2 \\ p_2, q_2\end{array};
\begin{array}{c}\alpha_3 \\ p_3, q_3\end{array};
\begin{array}{c}\alpha_4 \\ p_4, q_4\end{array}; \, \cdots \, ;
\begin{array}{c}\alpha_{2N-1} \\ p_{2N-1}, q_{2N-1}\end{array};
\begin{array}{c}\alpha_{2N} \\ p_{2N}, q_{2N}\end{array}\right ]}
\\ \\
\displaystyle{=\det_{1\leq i,k\leq 2N}\left ( \delta_{ik}+
\alpha_i \, \frac{p_i-q_i}{p_i-q_k}\, e^{(p_i-q_i)x+\xi ({\bf t}, p_i)-\xi ({\bf t}, q_i)}
\right ).}
\end{array}
\eeq
The $N$-soliton tau-function of the BKP hierarchy is the square root of the $\tau^{\rm KP}$ 
specialized as
\beq\label{ms2}
\begin{array}{c}
\displaystyle{
\tau^{\rm KP} \left [ 
\begin{array}{c}-q_1\alpha_1 \\ p_1, -q_1\end{array};
\begin{array}{c}p_1\alpha_1 \\ q_1, -p_1\end{array};
\begin{array}{c}-q_2\alpha_2 \\ p_2, -q_2\end{array};
\begin{array}{c}p_2\alpha_2 \\ q_2, -p_2\end{array}; \, \cdots \, ;
\begin{array}{c}-q_N\alpha_{N} \\ p_{N}, -q_{N}\end{array};
\begin{array}{c}p_N\alpha_{N} \\ q_{N}, -p_{N}\end{array}\right ]},
\end{array}
\eeq
where it is implied that even times evolution is frozen ($t_{2k}=0$ for all $k\geq 1$). 
It can be proved that this KP tau-function is a full square, i.e., 
$\tau =\sqrt{\vphantom{B^{a^a}}\tau^{\rm KP}}$ is an entire function of the times
(no square root singularities!).

\section{The CKP hierarchy}

In this section we present the main formulas
related to the CKP hierarchy with some details. 
The main references are \cite{DJKM81,DM-H09,CW13}.

\subsection{The CKP equation and the hierarchy}

The set of independent variables (``times'') is ${\bf t}_{\rm o}
=\{t_1, t_3, t_5, \ldots \}$.
Like in the BKP hierarchy, they are indexed by positive odd numbers. We set
$t_1 = x+\mbox{const}$.
The main object is the $L$-operator (\ref{kp1}) 
with the constraint
\beq\label{ckp2}
L^{\dag}=-L.
\eeq
It is instructive to reformulate this constraint in terms of the
wave operator $W$ (\ref{bkp1a}) such that $L=W\p_x W^{-1}$. 
The constraint implies that $W^{\dag}W$ commutes with $\p_x$, i.e., 
it is a pseudo-differential operator with constant coefficients. 
The freedom in the definition of the wave operator can be fixed by demanding that
$W^{\dag}W=1$, i.e.
\beq\label{ckp1b}
W^{\dag}=W^{-1}.
\eeq
The evolution equations (the Lax equations) 
and the zero curvature equations have the same form (\ref{bkp3}) 
and (\ref{bkp3a}) as in the BKP hierarchy.
By applying
the $\dag$-operation to the evolution equations (\ref{bkp3})
it is not difficult to see that they are consistent with the
constraint (\ref{ckp2}), i.e., 
$\p_{t_k}(L+L^{\dag})=0$ for odd $k$, so the CKP hierarchy is well-defined.

Clearly, the differential operators $B_k$ satisfy $B_k^{\dag}=-B_k$ (for odd $k$).
In particular,
\beq\label{ckp5}
\begin{array}{l}
B_1=\p_x,
\\ \\
B_3 =\p_x^3 +6u\p_x +3u', 
\\ \\
B_5 =\p_x^5 +10u\p^3_x +15u' \p_x^2 +v\p_x +\frac{1}{2}\, (v'-5u''').
\end{array}
\eeq
where $u'\equiv \p_x u$, $u=\frac{1}{2}\, u_1$.
Since $B_1=\p_x$, it follows from (\ref{bkp3}), like in the KP hierarchy, 
that the evolution in the time
$t_1$ is simply the shift of $x$, i.e. the solutions depend on $x+t_1$.

The first equation of the CKP hierarchy follows from the zero curvature
equation $\p_{t_3}B_5-\p_{t_5}B_3+[B_5, B_3]=0$ with $B_3$, $B_5$ 
as in (\ref{ckp5}). The calculations yield the following
system of equations for the unknown functions $u,v$:
\beq\label{ckp0}
\left \{ \begin{array}{l}
10\p_{t_3}u=3v' -35u''' -120 uu'
\\ \\
6\p_{t_5}u-\p_{t_3}v=\frac{57}{2}\, u''''' +150 uu''' +180u'u'' -
\frac{5}{2}\, v''' +6vu'-6uv'.
\end{array} \right.
\eeq
Note that the variable $v$ can be excluded by passing to the unknown function $U$
such that $U'=u$.

\subsection{The wave function and the tau-function}

Like in the KP and BKP hierarchies, the Lax equations and the zero curvature equations
are compatibility conditions of the auxiliary linear problems
\beq\label{ckp4}
\p_{t_k}\Psi =B_k \Psi , \quad L\Psi =z\Psi
\eeq
for the formal wave function
\beq\label{ckp4a}
\Psi=\Psi(x, {\bf t}_{\rm o}, z)=We^{xz+\xi ({\bf t}_{\rm o}, z)},
\eeq
where $z$ is the spectral parameter and $\xi ({\bf t}_{\rm o}, z)$ 
is defined in (\ref{bkp7a}). 
The wave function has the following expansion as $z\to \infty$:
\beq\label{ckp5b}
\Psi(x, {\bf t}_{\rm o}, z)=e^{xz+\xi ({\bf t}_{\rm o}, z)} 
\Bigl (1+\sum_{k\geq 1}\xi_k z^{-k}\Bigr ).
\eeq
As is proved in \cite{DJKM81}, it satisfies the
bilinear relation
\beq\label{ckp5a}
\oint_{C_{\infty}}\! \Psi (x, {\bf t}_{\rm o}, z)
\Psi (x, {\bf t}_{\rm o}', -z)\frac{dz}{2\pi i}=0
\eeq
valid for all ${\bf t}_{\rm o}, {\bf t}_{\rm o}'$.
Here is a sketch of proof.
By virtue of the differential equations (\ref{ckp4}), the bilinear relation is equivalent
to vanishing of
$$
b_m=\p_{x'}^m \oint_{C_{\infty}}\! \Psi (x, {\bf t}_{\rm o}, z)\Psi (x', {\bf t}_{\rm o}, -z)
\frac{dz}{2\pi i} \Biggr |_{x'=x}\quad \mbox{for all $m\geq 0$.}
$$
The further calculation is similar to the one done in the BKP case. 
We have:
$$
b_m=\oint_{C_{\infty}}\Bigl (\sum_{k\geq 0}\xi_k(x)z^{-k}\Bigr )\p_{x'}^m
\Bigl (\sum_{l\geq 0}\xi_l(x')(-z)^{-l}\Bigr )e^{(x-x')z} \frac{dz}{2\pi i} \Biggr |_{x'=x}
$$
$$
=\oint_{C_{\infty}}\Bigl (\sum_{k\geq 0}\xi_kz^{-k}\Bigr )(\p_x -z)^m
\Bigl (\sum_{l\geq 0}\xi_l(-z)^{-l}\Bigr )\frac{dz}{2\pi i}
$$
$$
=\sum_{j+k+l=m+1}(-1)^{m+j+l}\left (\! \begin{array}{c}m\\ j \end{array} \! \right )
\xi_k \p_x^j \xi_l.
$$
It remains to notice that this 
expression is the coefficient of $(-1)^m \p_x^{-m-1}$ in the operator
$WW^{\dag}$:
$\displaystyle{
WW^{\dag}=1+\sum_{m\geq 0}(-1)^m b_m \p_x^{-m-1}}.
$
Since $WW^{\dag}=1$ (see (\ref{ckp1b})), we conclude that $b_m=0$ for all $m\geq 0$.

The tau-function $\tau = \tau (x, {\bf t}_{\rm o})$
of the CKP hierarchy is consistently introduced by the formula \cite{DM-H09,CW13}
\beq\label{ckp6}
\Psi  = e^{xz+\xi ({\bf t}_{\rm o}, z)}G(x, {\bf t}_{\rm o}, z)\,
\frac{\tau (x, {\bf t}_{\rm o}-2[z^{-1}]_{\rm o})}{\tau (x, {\bf t}_{\rm o})},
\eeq
where
\beq\label{ckp6a}
G(x, {\bf t}_{\rm o}, z)=\left (1+z^{-1}\p_{t_1}\log 
\frac{\tau (x, {\bf t}_{\rm o}-2[z^{-1}]_{\rm o})}{\tau (x,{\bf t}_{\rm o})}
\right )^{1/2}
\eeq
(cf. (\ref{bkp6}), where there is no factor $G(x, {\bf t}_{\rm o}, z)$).
The formula (\ref{ckp6}) (and the very existence of the tau-function) 
is based on the bilinear relation (\ref{ckp5a}) The proof can be found in \cite{KZ20}.
For completeness, we give it here.
Let us represent the wave function
in the form
$$
\Psi (x, {\bf t}_{\rm o}, z)=e^{xz+\xi ({\bf t}_{\rm o}, z)}w(x,{\bf t}_{\rm o}, z)
$$
and set ${\bf t}_{\rm o}'={\bf t}_{\rm o}-2[a^{-1}]_{\rm o}$ in the bilinear relation. 
We have
$\displaystyle{e^{\xi ({\bf t}_{\rm o}-{\bf t}_{\rm o}', z)}=
\frac{a+z}{a-z}}$. The residue calculus in (\ref{ckp5a})
yields
\beq\label{der1}
w({\bf t}_{\rm o}, a) w({\bf t}_{\rm o}\! -\! 2[a^{-1}]_{\rm o}, -a)=f({\bf t}, a),
\eeq
where
\beq\label{der2}
f({\bf t}_{\rm o}, z)=1+\frac{1}{2z}\, \Bigl (\xi_1({\bf t}_{\rm o})-
\xi_1({\bf t}_{\rm o}\! -\! 2[z^{-1}]_{\rm o})\Bigr )
\eeq
(here and below we do not indicate the $x$-dependence for brevity). 
Next, we set ${\bf t}_{\rm o}'={\bf t}_{\rm o}-2[a^{-1}]_{\rm o}-2[b^{-1}]_{\rm o}$
in the bilinear relation and the residue calculus yields
\beq\label{der7}
\begin{array}{c}
\displaystyle{\frac{a+b}{a-b}\Bigl (aw({\bf t}_{\rm o}, a)w({\bf t}_{\rm o}\! -\!
2[a^{-1}]_{\rm o}\!
-\! 2[b^{-1}]_{\rm o}, -a)-
bw({\bf t}_{\rm o}, b)w({\bf t}_{\rm o}\! -\! 2[a^{-1}]_{\rm o}
\! -\! 2[b^{-1}]_{\rm o}, -b)\Bigr )}
\\ \\
\displaystyle{=
a+b+\frac{1}{2}\Bigl (\xi_1({\bf t}_{\rm o})-\xi_1({\bf t}_{\rm o}\! -\! 2[a^{-1}]_{\rm o}
\! -\! 2[b^{-1}]_{\rm o}\Bigr )}.
\end{array}
\eeq
Expressing $w(\ldots , -a)$, $w(\ldots , -b)$ through $w(\ldots , a)$, $w(\ldots , b)$
by means of the relation (\ref{der1}), we can represent this equation in the form
\beq\label{der8}
\begin{array}{c}
\displaystyle{
\frac{1}{a-b}\left (af({\bf t}_{\rm o}\! -\! 2[b^{-1}]_{\rm o}, a)
\frac{w({\bf t}_{\rm o}, a)}{w({\bf t}_{\rm o}\! -\! 2[b^{-1}]_{\rm o},a)}-
bf({\bf t}_{\rm o}\! -\! 2[a^{-1}]_{\rm o}, b)
\frac{w({\bf t}_{\rm o}, b)}{w({\bf t}_{\rm o}\! -\! 2[a^{-1}]_{\rm o},b)}\right )
}
\\ \\
\displaystyle{=\, 1+\frac{\xi_1({\bf t}_{\rm o})-
\xi_1({\bf t}_{\rm o}\! -\! 2[a^{-1}]_{\rm o}\! -\! 2[b^{-1}]_{\rm o})}{2(a+b)}.}
\end{array}
\eeq
Shifting here ${\bf t}_{\rm o}\rightarrow {\bf t}_{\rm o}+2[b^{-1}]_{\rm o}$,
changing the sign of $b$ (i.e, substituting $b\to -b$)
and using (\ref{der1}) again after that (in the second term in the
left hand side), we arrive at the equivalent equation
\beq\label{der9}
\begin{array}{c}
\displaystyle{
\frac{1}{a+b}\left (af({\bf t}_{\rm o}, a)
\frac{w({\bf t}_{\rm o}\! -\! 2[b^{-1}]_{\rm o}, a)}{w({\bf t}_{\rm o},a)}-
bf({\bf t}_{\rm o}, b)
\frac{w({\bf t}_{\rm o}\! -\! 2[a^{-1}]_{\rm o}, b)}{w({\bf t}_{\rm o},b)}\right )
}
\\ \\
\displaystyle{=\, 1+\frac{\xi_1({\bf t}_{\rm o}\! -\! 2[b^{-1}]_{\rm o})-
\xi_1({\bf t}_{\rm o}\! -\! 2[a^{-1}]_{\rm o})}{2(a-b)}.}
\end{array}
\eeq
Together equations (\ref{der8}), (\ref{der9}) form the system of two equations
\beq\label{der10}
\left \{
\begin{array}{l}
\displaystyle{\frac{1}{a\! -\! b}\left (af({\bf t}_{\rm o}\! -\! 2[b^{-1}]_{\rm o}, a)X^{-1}\! -\!
bf({\bf t}_{\rm o}\! -\! 2[a^{-1}]_{\rm o}, b)Y^{-1}\right )\! =\!
\frac{af({\bf t}_{\rm o}, a)\! +\! bf({\bf t}_{\rm o}\! -\! 2[a^{-1}]_{\rm o}, b)}{a+b}}
\\ \\
\displaystyle{\frac{1}{a\! +\! b}\, \Bigl (af({\bf t}_{\rm o}, a)X -
bf({\bf t}_{\rm o}, b)Y\Bigr ) =
\frac{af({\bf t}_{\rm o}, a) - bf({\bf t}_{\rm o}, b)}{a-b}}
\end{array}
\right.
\eeq
for the two ``unknown quantities''
\beq\label{der11}
X=\frac{w({\bf t}_{\rm o}-2[b^{-1}]_{\rm o}, a)}{w({\bf t}_{\rm o}, a)}, \qquad
Y=\frac{w({\bf t}_{\rm o}-2[a^{-1}]_{\rm o}, b)}{w({\bf t}_{\rm o}, b)}.
\eeq
The next step is to multiply the two equations (\ref{der10}). 
After some algebra, one obtains the following simple relation:
\beq\label{der13}
\frac{Y}{X}=\frac{w({\bf t}_{\rm o}, a)
w({\bf t}_{\rm o}-2[a^{-1}]_{\rm o}, b)}{w({\bf t}_{\rm o}, b)
w({\bf t}_{\rm o}-2[b^{-1}]_{\rm o}, a)}=
\left (\frac{f({\bf t}_{\rm o}, a)
f({\bf t}_{\rm o}-2[a^{-1}]_{\rm o}, b)}{f({\bf t}_{\rm o}, b)
f({\bf t}_{\rm o}-2[b^{-1}]_{\rm o}, a)}\right )^{1/2}.
\eeq
In the calculations, the identity
\beq\label{der12}
af({\bf t}_{\rm o}, a)-af({\bf t}_{\rm o}-2[b^{-1}]_{\rm o}, a)
-bf({\bf t}_{\rm o}, b)+bf({\bf t}_{\rm o}-2[a^{-1}]_{\rm o}, b)=0
\eeq
has been used.
Introducing $w_0({\bf t}_{\rm o}, z)=w({\bf t}_{\rm o}, z)
f^{-1/2}({\bf t}_{\rm o}, z)$, we get from it the relation
\beq\label{der14}
\frac{w_0({\bf t}_{\rm o}, a)
w_0({\bf t}_{\rm o}-2[a^{-1}]_{\rm o}, b)}{w_0({\bf t}_{\rm o}, b)
w_0({\bf t}_{\rm o}-2[b^{-1}]_{\rm o}, a)}=1
\eeq
which has the same form as (\ref{bder8}) for the BKP hierarchy, with the change
in the notation $w\to w_0$. 

As soon as the relation of the form (\ref{der14}) is established, 
the rest of the argument is the same as for the BKP hierarchy. 
In the same way as in section \ref{section:wavetau} we can
prove that there exists a function $\tau ({\bf t}_{\rm o})$ such that
\beq\label{der15}
w_0({\bf t}_{\rm o}, z)=\frac{\tau ({\bf t}_{\rm o}-2[z^{-1}]_{\rm o})}{\tau ({\bf t}_{\rm o})}.
\eeq
This function is called the tau-function of the CKP hierarchy. 
Finally, writing $w({\bf t}_{\rm o}, z)=f^{1/2}({\bf t}_{\rm o}, z)w_0({\bf t}_{\rm o}, z)$
and noting that $f({\bf t}_{\rm o}, z)=1+O(z^{-2})$, we see that
\beq\label{der18}
\xi_1 ({\bf t}_{\rm o})=-2\p_{t_1}\log \tau ({\bf t}_{\rm o})
\eeq
and, recalling (\ref{der2}), we arrive at (\ref{ckp6})
with $G=f^{1/2}$.

Let us show that equation (\ref{ckp6}) can be obtained up to a common
$x$-independent factor in the following easy way \cite{DM-H09,CW13}.
Apply $\p_{t_1}$ to
(\ref{ckp5a}) and set ${\bf t}_{\rm o}'={\bf t}_{\rm o}-2[a^{-1}]_{\rm o}$.
The residue calculus yields
\beq\label{der3}
\begin{array}{c}
2a^2\Bigl (1\! -\! w({\bf t}_{\rm o}, a) w({\bf t}_{\rm o}\! -\! 2[a^{-1}]_{\rm o}, -a)\Bigr )-2a
w'({\bf t}_{\rm o}, a) w({\bf t}_{\rm o}-2[a^{-1}]_{\rm o}, -a)
\\ \\
+2a\Bigl (\xi_1({\bf t}_{\rm o})-\xi_1({\bf t}_{\rm o}\! -\!
2[a^{-1}]_{\rm o})\Bigr ) +\xi_2({\bf t}_{\rm o}\! -\! 2[a^{-1}]_{\rm o})
+\xi_2({\bf t}_{\rm o})+\xi_1'({\bf t}_{\rm o})
\\ \\
\phantom{aaaaaaaaaaaaaaaaaaaaaaaa}-
\xi_1({\bf t}_{\rm o})\xi_1({\bf t}_{\rm o}\! -\! 2[a^{-1}]_{\rm o})=0,
\end{array}
\eeq
where prime means the $x$-derivative and
we again do not indicate the dependence on $x$ explicitly.
Tending $a\to \infty$ in (\ref{der3}), we get the relation
\beq\label{der4}
2\xi_2 ({\bf t}_{\rm o})=\xi_1^2 ({\bf t}_{\rm o})-\xi_1' ({\bf t}_{\rm o})
\eeq
(it also directly follows from the constraint $WW^{\dag}=1$ for the dressing operator).
Plugging it back to (\ref{der3}), we can rewrite this equation in the form
\beq\label{der5}
\begin{array}{c}
w'({\bf t}_{\rm o}, a) w({\bf t}_{\rm o}-2[a^{-1}]_{\rm o}, -a)=
af({\bf t}_{\rm o}, a)(f({\bf t}_{\rm o}, a)-1)+\frac{1}{2}
f'({\bf t}_{\rm o}, a).
\end{array}
\eeq
Using (\ref{der1}), we conclude that
\beq\label{der6}
\begin{array}{c}
\p_x \log w({\bf t}_{\rm o}, a)=a(f({\bf t}_{\rm o}, a)-1)+\frac{1}{2}\, \p_x
\log f({\bf t}_{\rm o}, a)
\\ \\
=\frac{1}{2}\Bigl (\xi_1({\bf t}_{\rm o})-\xi_1({\bf t}_{\rm o}-2[a^{-1}]_{\rm o})\Bigr )+
\frac{1}{2}\, \p_x \log f({\bf t}_{\rm o}, a).
\end{array}
\eeq
Now, setting $\xi_1(x, {\bf t}_{\rm o})=-2\p_x\log \tau (x, {\bf t}_{\rm o})$ 
with some function $\tau (x, {\bf t}_{\rm o})$
and integrating,
we arrive at (\ref{ckp6})
with $G=f^{1/2}$ up to a common multiplier which does not depend on $x$.

The function $u$ in (\ref{ckp5}) can be also expressed
through the tau-function.
It is a matter of direct verification that the result of the action of the operator
$\p^3_x +6u\p_x +3u' -\p_{t_3}$ to the wave function $\Psi$ of the form (\ref{ckp5b}) is
$O(z^{-1})$ as $z\to \infty$, i.e.,
\beq\label{ckp5e}
(\p^3_x +6u\p_x +3u' -\p_{t_3})\Psi =O(z^{-1})e^{xz+\xi ({\bf t}_{\rm o}, z)}
\eeq
if the conditions
\beq\label{ckp5d}
\begin{array}{c}
u=-\frac{1}{2}\, \xi_1', \qquad u'=\xi_1\xi_1' -\xi_1'' -\xi_2'
\end{array}
\eeq
hold true. Since from (\ref{ckp6}) it follows that
$$\xi_1=-2\p_x\log \tau , \quad \xi_2=2(\p_x \log \tau )^2 +\p_x^2\log \tau ,$$ we have
\beq\label{ckp10}
u=\p_x^2\log \tau
\eeq
and the second equality in (\ref{ckp5d}) holds identically.

\subsection{The equation for the tau-function}

As it follows from (\ref{ckp5a}), (\ref{ckp6}),
the CKP hierarchy is equivalent to the following relation for the
tau-function:
\beq\label{ckp9}
\oint_{C_{\infty}} \!\! e^{\xi ({\bf t}_{\rm o}-{\bf t}_{\rm o}', z)}
\tau \Bigl (x,{\bf t}_{\rm o}-2[z^{-1}]_{\rm o}\Bigr )
\tau \Bigl (x, {\bf t}_{\rm o}'+2[z^{-1}]_{\rm o}\Bigr )
G(x, {\bf t}_{\rm o}, z)G(x, {\bf t}_{\rm o}', -z)\, \frac{dz}{2\pi i} =0
\eeq
valid for all ${\bf t}_{\rm o}, {\bf t}_{\rm o}'$.  Set ${\bf t}'_{\rm o}=
{\bf t}_{\rm o}-2[a^{-1}]_{\rm o}-
2[b^{-1}]_{\rm o}-2[c^{-1}]_{\rm o}$, then
$$
e^{\xi ({\bf t}_{\rm o}-{\bf t}_{\rm o}', z)}=\frac{(a+z)(b+z)(c+z)}{(a-z)(b-z)(c-z)}
$$
and the residue calculus in (\ref{ckp9}) gives the following equation:
\beq\label{ckp11}
\begin{array}{c}
a(a+b)(a+c)(b-c)\tau^{[a]_{\rm o}}\tau^{[bc]_{\rm o}}G(-a)G^{[abc]_{\rm o}}(a)
\\ \\
+ b(b+a)(b+c)(c-a)\tau^{[b]_{\rm o}}\tau^{[ac]_{\rm o}}G(-b)G^{[abc]_{\rm o}}(b)
\\ \\
+ c(c+a)(c+b)(a-b)\tau^{[c]_{\rm o}}\tau^{[ab]_{\rm o}}G(-c)G^{[abc]_{\rm o}}(c)
\\ \\
+(a+b+c)(a-b)(b-c)(c-a)\tau \tau^{[abc]_{\rm o}}
\\ \\
+(a-b)(b-c)(c-a)\Bigl (\tau^{[abc]_{\rm o}}\p_{t_1}\tau - \tau \p_{t_1}\tau^{[abc]_{\rm o}}\Bigr )
=0,
\end{array}
\eeq
where
$
\tau^{[a]_{\rm o}}=\tau (x, {\bf t}_{\rm o}+2[a^{-1}]_{\rm o})$,
$\tau^{[ab]_{\rm o}}=\tau (x, {\bf t}_{\rm o}+2[a^{-1}]_{\rm o}+2[b^{-1}]_{\rm o})$,
$G^{[a]_{\rm o}}(z)=G(x, {\bf t}_{\rm o}+2[a^{-1}]_{\rm o}, z)$
and so on.
Equation (\ref{ckp11}) should be valid for all $a,b,c$.
Let us tend $c$ to infinity. The highest terms
proportional to $c^3$ vanish identically.
The terms of order $c^2$ give the equation
\beq\label{ckp12}
\frac{a+b}{a-b}\Bigl (aG(-a)G^{[ab]_{\rm o}}(a)-bG(-b)G^{[ab]_{\rm o}}(b)\Bigr )=
\left (a+b -\p_{t_1}\log \frac{\tau^{[ab]_{\rm o}}}{\tau}\right )
\frac{\tau \tau^{[ab]_{\rm o}}}{\tau^{[a]_{\rm o}}\tau^{[b]_{\rm o}}},
\eeq
or, in the more detailed notation,
\beq\label{ckp13}
\begin{array}{c}
\displaystyle{\frac{z_1+z_2}{z_1-z_2}\left [
\Bigl (z_1-\p_{t_1}\log \frac{\tau^{[z_1z_2]_{\rm o}}}{\tau^{[z_2]_{\rm o}}}\Bigr )^{1/2}
\Bigl (z_1-\p_{t_1}\log \frac{\tau^{[z_1]_{\rm o}}}{\tau}\Bigr )^{1/2}\right. }
\\ \\
\phantom{aaaaaaaaaaaaaaaaaaaaaaaaa}\displaystyle{\left.
-\Bigl (z_2-\p_{t_1}\log \frac{\tau^{[z_1z_2]_{\rm o}}}{\tau^{[z_1]_{\rm o}}}\Bigr )^{1/2}
\Bigl (z_2-\p_{t_1}\log \frac{\tau^{[z_2]_{\rm o}}}{\tau}\Bigr )^{1/2}\right ]}
\\ \\
=\displaystyle{ \left (z_1+z_2 -\p_{t_1}\log \frac{\tau^{[z_1z_2]_{\rm o}}}{\tau}\right )
\frac{\tau \, \tau^{[z_1z_2]_{\rm o}}}{\tau^{[z_1]_{\rm o}}\tau^{[z_2]_{\rm o}}}.}
\end{array}
\eeq
This is the equation for the tau-function of the CKP hierarchy. It should hold for all
$z_1, z_2$. In contrast to the cases of the KP and BKP hierarchies, this equation
is not bilinear.

\subsection{The CKP hierarchy and the KP hierarchy}

Here we give a characterization of those KP tau-functions which
correspond to solutions of the CKP hierarchy and show that the CKP tau-function
is the square root of the KP one. 

Comparing (\ref{ch2}) and (\ref{ckp5a}), we conclude that the wave functions of the CKP and KP
hierarchies are related as
$$
\begin{array}{l}
\Psi (x, {\bf t}_{\rm o}, z)=e^{\chi (z)}\Psi^{\rm KP}(x, t_1, 0, t_3, 0, \ldots , z),
\\ \\
\Psi (x, {\bf t}_{\rm o}, -z)=e^{-\chi (z)}
\Psi^{\dag \rm KP}(x, t_1, 0, t_3, 0, \ldots , z).
\end{array}
$$
Here $\chi (z)$ is some function such that $\chi (\infty )=0$, i.e.
\beq\label{ch4a}
\Psi^{\dag \rm KP}(x, t_1, 0, t_3, 0, \ldots , z)=
e^{2\chi_{\rm e}(z)}\Psi^{\rm KP}(x, t_1, 0, t_3, 0, \ldots , -z),
\eeq
where $\chi_{\rm e}(z)=\frac{1}{2}(\chi(z)+\chi (-z))$ is the even part of the function $\chi (z)$.
From (\ref{ch1}), (\ref{ch1a}) and (\ref{ch4a}) it follows
that the KP tau-function is the extension of a solution of the CKP hierarchy
if and only if the equation
\beq\label{ch4}
\begin{array}{l}
\tau^{\rm KP}\Bigl (x, t_1+z^{-1}, \frac{1}{2}\, z^{-2},
t_3+\frac{1}{3}\, z^{-3}, \frac{1}{4}\, z^{-4}, \ldots \Bigr )
\\ \\
\phantom{aaaaaaaaaaa}=e^{2\chi_{\rm e}(z)}\tau^{\rm KP}\Bigl (x, t_1+z^{-1}, -\frac{1}{2}\, z^{-2},
t_3+\frac{1}{3}\, z^{-3}, -\frac{1}{4}\, z^{-4}, \ldots \Bigr )
\end{array}
\eeq
holds identically for all $z, x, t_1, t_3, t_5, \ldots$. Shifting the odd times, we can
rewrite this condition as
\beq\label{ch4b}
\begin{array}{c}
\log \tau^{\rm KP}\Bigl (x, t_1, \frac{1}{2}\, z^{-2},
t_3, \frac{1}{4}\, z^{-4}, \ldots \Bigr )
-\log \tau^{\rm KP}\Bigl (x, t_1, -\frac{1}{2}\, z^{-2},
t_3, -\frac{1}{4}\, z^{-4}, \ldots \Bigr )=2\chi_{\rm e}(z).
\end{array}
\eeq
Comparing the coefficients at $z^{-2}$ of the expansions of the 
left and right hand sides of (\ref{ch4b})
and passing to an equivalent tau-function
if necessary, we get the condition
\beq\label{ch4c}
\p_{t_2}\log \tau^{\rm KP} \Bigr |_{t_{2k}=0}=0, \quad k\geq 1.
\eeq
It is the CKP counterpart of the condition (\ref{comp7}) specific for the BKP
hierarchy.
In \cite{KZ20} it is proved that this is also a sufficient condition that the 
tau-function $\tau^{\rm KP}$ generates a solution to the CKP hierarchy.
The proof is based on the technique developed in \cite{Natanzon1,Natanzon2,NZ16}.

Let us introduce the auxiliary wave function $\psi$ in the same way as for the BKP
hierarchy:
\beq\label{e3}
\psi = e^{xz+\xi ({\bf t}_{\rm o}, z)}
\frac{\tau (x, {\bf t}_{\rm o}-2[z^{-1}]_{\rm o})}{\tau (x, {\bf t}_{\rm o})},
\eeq
then the wave function (\ref{ckp6}) is
\beq\label{e4}
\Psi =z^{-1/2} \sqrt{\vphantom{B^{a^a}}\p_x \log \psi} \cdot \psi =
(2z)^{-1/2}\sqrt{\vphantom{B^{a^a}}\p_x \psi^2}.
\eeq
We will prove that the CKP and KP tau-functions 
are related as
\beq\label{e1}
\tau =\sqrt{\vphantom{B^{a^a}}\tau^{\rm KP}}.
\eeq
The tau-function $\tau (x, {\bf t}_{\rm o})$ has square root singularities in $x$. However,
it appears that the expression $\p_x \psi^2$
is a full square for all $z$, i.e., $\sqrt{\vphantom{B^{a^a}}\p_x \psi^2}$ and, therefore,
$\Psi$ is an entire
function of $x$ and ${\bf t}_{\rm o}$ (no square root singularities!). This is similar to the
fact that for the BKP hierarchy the function $\tau^{\rm KP}$ is a full square.
The bilinear identity (\ref{ckp5a}) acquires the form
\beq\label{ckp5c}
\oint_{C_{\infty}} \sqrt{\vphantom{B^{a^a}}\p_x \psi^2 (x, {\bf t}_{\rm o}, z)}\,
\sqrt{\vphantom{B^{a^a}}\p_x \psi^2 (x, {\bf t}_{\rm o}', -z)}\, \frac{dz}{2\pi i z}=0.
\eeq

In order to prove that $\tau =\sqrt{\vphantom{A^A}\tau^{\rm KP}}$  (see \cite{CH14})
we compare two expressions for the wave
function $\Psi$ of the CKP hierarchy. The first one is
in terms of the KP tau-function,
\beq\label{e5a}
\Psi^{\rm KP}
= e^{xz+\xi ({\bf t}_{\rm o}, z)}\frac{\tau^{\rm KP}(x, t_1-z^{-1}, -\frac{1}{2}\, z^{-2},
t_3-\frac{1}{3}\, z^{-3}, -\frac{1}{4}\, z^{-4}, \ldots )}{\tau^{\rm KP}
(t_1, 0, t_3, 0, \ldots )},
\eeq
and the second one (\ref{ckp6}) is in terms of the CKP tau-function $\tau$.
Comparing (\ref{e4})
and (\ref{e5a}), we get the equation
\beq\label{e5b}
\frac{1}{2z}\, \p_x \!\! \left (e^{2xz}\frac{\tau^2 \Bigl (
x, {\bf t}_{\rm o}-2[z^{-1}]_{\rm o}\Bigr )}{\tau^2 (
x, {\bf t}_{\rm o})}\right )=
e^{2xz}\left (\frac{\tau^{\rm KP}(x, {\bf \dot t}-[z^{-1}])}{\tau^{\rm KP}
(x, {\bf \dot t})}\right )^2,
\eeq
where we again use the short-hand notation ${\bf \dot t}=\{t_1, 0, t_3, 0, \ldots \}$.
Then
we get that (\ref{e5b}) is equivalent to
the differential equation
$
\p_x \varphi = -2z \varphi ,
$
where
\beq\label{e7}
\varphi = \frac{\tau^2 \Bigl (x, {\bf t}_{\rm o}-2[z^{-1}]_{\rm o}
\Bigr )}{\tau^2 (x, {\bf t}_{\rm o})}-
\frac{\tau^{\rm KP}\Bigl (x, {\bf \dot t}-2[z^{-1}]_{\rm o}\Bigr )}{\tau^{\rm KP}
(x, {\bf \dot t})}.
\eeq
In (\ref{e7}), ${\bf \dot t}-2[z^{-1}]_{\rm o}=\{t_1-2z^{-1}, 0, t_3-\frac{2}{3}\, z^{-3}, 0,
\ldots \}$.
The general solution of the differential equation is
$
\varphi =c(z, t_3, t_5, \ldots )e^{-2(x+t_1)z}
$
but from (\ref{e7}) it follows that $\varphi$
is expanded in a power series
as $\varphi =\varphi_1 z^{-1}+\varphi_2 z^{-2}+\ldots$ as $z\to \infty$, and
this means that $c$ must be equal to 0. Therefore, $\varphi=0$, i.e.
\beq\label{e8}
\frac{\tau^2 \Bigl (x, {\bf t}_{\rm o}-2[z^{-1}]_{\rm o}
\Bigr )}{\tau^2 (x, {\bf t}_{\rm o})}=
\frac{\tau^{\rm KP}\Bigl (x, {\bf \dot t}-2[z^{-1}]_{\rm o}\Bigr )}{\tau^{\rm KP}
(x, {\bf \dot t})}
\eeq
for all $z$. This is an identity on solutions to the KP/CKP hierarchies. It follows from
(\ref{e8}) that $\tau^{\rm KP}={\rm const}\cdot \tau^2$, i.e.
$\tau (x, {\bf t}_{\rm o})=\sqrt{\tau^{\rm KP} (x, {\bf \dot t})}$ is a tau-function
of the CKP hierarchy.

\subsection{Examples: soliton solutions}

\subsubsection{One-soliton solution}

We start from the simplest example of one-soliton solution.
The tau-function for one CKP soliton is the square root
of a specialization of 2-soliton tau-function
of the KP hierarchy. The latter has the form
\beq\label{s1}
\begin{array}{l}
\displaystyle{
\tau^{\rm KP}(x, {\bf t})=1+\alpha \exp \Bigl (
(p+q)x +\sum_{k\geq 1}(p^k -(-q)^k)t_k \Bigr )}
\\ \\
\phantom{aaaaaaaaaaaaaaaaaaa}\displaystyle{
+\, \alpha \exp \Bigl (
(p+q)x +\sum_{k\geq 1}(q^k -(-p)^k)t_k \Bigr )}
\\ \\
\displaystyle{
\phantom{aaaaaaa}-\frac{\alpha^2 (p-q)^2}{4pq}\, \exp \Bigl (2(p+q)x +\sum_{k\geq 1}
(p^k +q^k -(-p)^k -(-q)^k)t_k\Bigr )},
\end{array}
\eeq
where $\alpha , p, q$ are arbitrary parameters.
When all even times are put equal to zero, $t_2=t_4=\ldots =0$, this expression
simplifies:
\beq\label{e2}
\tau^{\rm KP}= 1+2\alpha w-
\frac{\alpha^2 (p-q)^2}{4pq} \, w^2,
\eeq
where
\beq\label{e5}
w=e^{(p+q)x+ \xi ({\bf t}_{\rm o}, p)+\xi ({\bf t}_{\rm o}, q)}.
\eeq
We see that the tau-function $\tau = \sqrt{\tau^{\rm KP}}$ has
two square root singularities at the
points $w=w_{\pm}$,
$$
w_{\pm}=\pm \frac{2\sqrt{\vphantom{B}pq}}{\alpha (\! \sqrt{p}\, \mp \! \sqrt{q}\, )^2}.
$$

A direct calculation shows that $\p_x \psi^2$ (where $\psi$
is given by (\ref{e3})) for the solution (\ref{e2})
is a full square for all $z$:
$$
\frac{\p_x \psi^2}{2z}=e^{2xz+2\xi ({\bf t}_{\rm o}, z)}
\left (\frac{1+\displaystyle{\frac{\alpha (2z^2-p^2-q^2)}{(z+p)(z+q)}}\, w -
\displaystyle{\frac{\alpha^2 (p-q)^2}{4pq}}\,
\displaystyle{\frac{(z-p)(z-q)}{(z+p)(z+q)}\, w^2}}{1+2\alpha w-
\displaystyle{\frac{\alpha^2 (p-q)^2}{4pq}} \, w^2}\right )^2
$$
$$
\hspace{-2cm}
=\Bigl (\Psi^{\rm KP}(x, t_1, 0, t_3, 0, \ldots , z)\Bigr )^2,
$$
where $\Psi^{\rm KP}$ is constructed from the KP tau-function
$\tau^{\rm KP}$ (\ref{s1}) according to formula (\ref{ch1}).
Hence $(2z)^{-1/2}\sqrt{\vphantom{B^{a^a}}\p_x \psi^2}$ and, therefore,
$\Psi$ does not have square root singularities in the variable $z$.
The bilinear identity
(\ref{ckp5a}), which in the present case has the explicit form
$$
\oint_{C_{\infty}} e^{\xi ({\bf t}_{\rm o}, z)-\xi ({\bf t}_{\rm o}', z)}
\left (1+\frac{\alpha (2z^2-p^2-q^2)}{(z+p)(z+q)}\, w -
\frac{\alpha^2 (p-q)^2}{4pq}\, \frac{(z-p)(z-q)}{(z+p)(z+q)}\, w^2\right )
$$
$$
\times \left (1+\frac{\alpha (2z^2-p^2-q^2)}{(z-p)(z-q)}\, w' -
\frac{\alpha^2 (p-q)^2}{4pq}\, \frac{(z+p)(z+q)}{(z-p)(z-q)}\, w'^2\right )\frac{dz}{2\pi i}=0,
$$
can be checked directly by the residue calculus.
(Here $w'=e^{(p+q)x+ \xi ({\bf t}_{\rm o}', p)+\xi ({\bf t}_{\rm o}', q)}$.)

\subsubsection{Multi-soliton solutions}

The general KP tau-function for $M$-soliton solution has $3M$ arbitrary parameters
$\alpha_i$, $p_i$, $q_i$ ($i=1, \ldots , M$) and is given by equation (\ref{ms1}).
We denote this tau-function as
$$
\tau^{\rm KP} \left [ \begin{array}{c}\alpha_1 \\ p_1, q_1\end{array};
\begin{array}{c}\alpha_2 \\ p_2, q_2\end{array};
\begin{array}{c}\alpha_3 \\ p_3, q_3\end{array};
\begin{array}{c}\alpha_4 \\ p_4, q_4\end{array}; \, \cdots \, ;
\begin{array}{c}\alpha_{2N-1} \\ p_{M-1}, q_{M-1}\end{array};
\begin{array}{c}\alpha_{2N} \\ p_{M}, q_{M}\end{array}\right ].
$$
The parameters $p_i, q_i$ are sometimes called momenta of solitons.

The multi-soliton tau-function of the CKP hierarchy is the square root of the $\tau^{\rm KP}$
specialized as
\beq\label{ms2a}
\begin{array}{c}
\displaystyle{
\tau^{\rm KP} \left [ \begin{array}{c}\alpha_0 \\ p_0, -p_0\end{array};
\begin{array}{c}\alpha_1 \\ p_1, -q_1\end{array};
\begin{array}{c}\alpha_1 \\ q_1, -p_1\end{array};
\begin{array}{c}\alpha_2 \\ p_2, -q_2\end{array};
\begin{array}{c}\alpha_2 \\ q_2, -p_2\end{array}; \, \cdots \, ;
\begin{array}{c}\alpha_{N} \\ p_{N}, -q_{N}\end{array};
\begin{array}{c}\alpha_{N} \\ q_{N}, -p_{N}\end{array}\right ]},
\end{array}
\eeq
where it is implied that even times evolution is suppressed ($t_{2k}=0$ for all $k\geq 1$).
Clearly, the total number of independent parameters is $3N+2$. If $\alpha_0=0$, the tau-function
(\ref{ms2}) reduces to
\beq\label{ms3a}
\begin{array}{c}
\displaystyle{
\tau^{\rm KP} \left [
\begin{array}{c}\alpha_1 \\ p_1, -q_1\end{array};
\begin{array}{c}\alpha_1 \\ q_1, -p_1\end{array};
\begin{array}{c}\alpha_2 \\ p_2, -q_2\end{array};
\begin{array}{c}\alpha_2 \\ q_2, -p_2\end{array}; \, \cdots \, ;
\begin{array}{c}\alpha_{N} \\ p_{N}, -q_{N}\end{array};
\begin{array}{c}\alpha_{N} \\ q_{N}, -p_{N}\end{array}\right ]},
\end{array}
\eeq
and it is this tau-function which is usually
called the $N$-soliton CKP tau-function in the literature (see, e.g. \cite{DJKM81}).
It is a specialization of $2N$-soliton KP tau-function and
has $3N$ free parameters.

\section{Dispersionless limit of the BKP and CKP hierarchies}

The dispersionless limit is the limit $\hbar \to 0$ after the substitutions
$$t_k\to t_k/\hbar, \qquad \tau =\exp (F/\hbar^2).$$
Making these substitutions in (\ref{bkp12}) and performing the limit, we get the equation
\beq\label{d1}
\frac{p(z_1)-p(z_2)}{z_1-z_2}=
\frac{p(z_1)+p(z_2)}{z_1+z_2}\, e^{4D^{\rm o}(z_1)D^{\rm o}(z_2)F},
\eeq
where
\beq\label{d3}
p(z)=z-2\p_{t_1}D^{\rm o}(z)F
\eeq
and 
$D^{\rm o}(z)$ is the differential operator
\beq\label{d2}
D^{\rm o}(z)=\sum_{k\geq 1, \, {\rm odd}}\frac{z^{-k}}{k}\, \p_{t_k}.
\eeq
The (odd) function $p(z)$ has the expansion
\beq\label{d8}
p(z)=z-\frac{u}{z}+\sum_{k\geq 3, \, {\rm odd}}u_kz^{-k},
\eeq
where
\beq\label{d9}
u=2\p_{t_1}^2 F.
\eeq
Equation (\ref{d1}) is the generating equation for the dispersionless BKP (dBKP)
hierarchy in the Hirota form \cite{Takasaki93,Takasaki06,Takebe14}. 
It is remarkable that the dispersionless limit of the CKP equation (\ref{ckp13})
is the same, so the dispersionless limits of the BKP and CKP hierarchies coincide.

Let us show how to represent the dispersionless hierarchy in the Lax form.
Taking logarithm of equation (\ref{d1}), differentiating with respect to $t_1$ and using
the definition (\ref{d8}), we obtain the equation
\beq\label{dbkp7}
2D^{\rm o}(z_1)p(z_2)=\p_{t_1} \log \frac{p(z_1)+p(z_2)}{p(z_1)-p(z_2)}
\eeq
from which it follows that $D^{\rm o}(z_1)p(z_2)=D^{\rm o}(z_2)p(z_1)$ (this follows also
from the definition (\ref{d8})). Tending $z_2 \to \infty$, we get
\beq\label{dbkp8}
\p_{t_1}p(z)=-D^{\rm o}(z)u.
\eeq
The next step is to 
rewrite equation (\ref{bkp7a}) in terms of the function $z(p)$, inverse to the $p(z)$
(like $p(z)$, it is an odd function with the Laurent series of the form 
$z(p)=p+ O(p^{-1})$).
Using the relation 
\beq\label{a14a}
\p_{t_k}p(z)=-\, \frac{\p_{t_k}z(p)}{\p_p z(p)}, \qquad k\geq 1,
\eeq 
we get, after simple transformations:
\beq\label{dbkp9}
2D^{\rm o}(z_1)z(p)=\left \{ z(p), \, \log \frac{p(z_1)-p}{p(z_1)+p}\right \},
\eeq
where 
\beq\label{a17a}
\{f,\, g\}:= \frac{\p f}{\p t_1}\, \frac{\p g}{\p p}- 
\frac{\p g}{\p t_1}\, \frac{\p f}{\p p}
\eeq
is the Poisson bracket.
This is the generating Lax equation
for the dBKP hierarchy, $z(p)$ being the Lax function
(the dispersionless limit of the Lax operator (\ref{bkp1})). 
Expanding equation (\ref{dbkp9}) in 
powers of $z_1$, one obtains the hierarchy of Lax equations through the Faber polynomials
${\cal B}_k(p)$ introduced by the expansion
\beq\label{a13}
-\log \frac{p(z)-p}{z} = \sum_{k\geq 1} \frac{z^{-k}}{k}\, {\cal B}_k(p).
\eeq
For example, ${\cal B}_1(p)=p$. It is easy to see that
\beq\label{a14}
{\cal B}_k(p)=\Bigl (z^k(p)\Bigr )_{\geq 0},
\eeq
where $(\ldots )_{\geq 0}$ is the polynomial part of the Laurent series 
in $p$ (containing only 
non-negative powers of the variable $p$). The fact that $p(z)$ is an odd function implies that
${\cal B}_k(-p)=(-1)^k {\cal B}_k(p)$ and we have the expansion
\beq\label{dbkp9a}
\log \frac{p(z)+p}{p(z)-p}=2\! \sum_{k\geq 1, \, {\rm odd}} \frac{z^{-k}}{k}\, {\cal B}_k(p).
\eeq
The Lax equations are of the form 
\beq\label{a18}
\p_{t_k}z(p)=\Bigl \{ {\cal B}_k(p), \, z(p)\Bigl \}=
\Bigl \{ (z^k(p))_{\geq 0}, \, z(p)\Bigl \}, \quad \mbox{$k$ odd}.
\eeq 

\section*{Acknowledgments}

\addcontentsline{toc}{section}{\hspace{6mm}Acknowledgments}

The author thanks V. Akhmedova, I. Krichever, S. Natanzon and D. Rudneva for discussions.
This work 
was supported by the Russian Science Foundation under grant  19-11-00275.

\end{document}